\documentclass[reprint,twocolumn]{revtex4}
\usepackage{amsfonts}
\usepackage{amssymb}
\usepackage{amsmath}
\usepackage{hyperref}
\usepackage{latexsym}
\usepackage{epsfig}

\begin{document}

\title{Cosmological wormholes }
\author{A.A. Kirillov}
\affiliation{Institute for Applied Mathematics and Cybernetics, 10
Uljanova Str., Nizhny Novgorod, 603005, Russia}
\author{E.P. Savelova}
\affiliation{Dubna International University of Nature, Society and
Man,  19 Universitetskaya Str., Dubna, 141980, Russia}

\begin{abstract}
We describe in details the procedure how the Lobachevsky space can be
factorized to a space of the constant negative curvature filled with a gas
of wormholes. We show that such wormholes have throat sections in the form
of tori and are traversable and stable in the cosmological context. The
relation of such wormholes to the dark matter phenomenon is briefly
described. We also discuss the possibility of the existence of analogous
factorizations for all types of homogeneous spaces.
\end{abstract}

\maketitle


\section{Introduction}

It was demonstrated by us previously that all features of cold dark matter
models (CDM) can be reproduced by the presence of a gas of wormholes \cite%
{KS11,KS07}. At very large scales wormholes behave like very heavy
particles, while at smaller subgalactic scales they strongly interact with
baryons and cure the problem of cusps. Lattice quantum gravity also suggests
strong theoretical arguments for fractal properties of the topological
structure of our Universe \cite{KPZ,AJW95,AJL05}. Such a structure has the
most natural realization as a homogeneous gas of wormholes with fractal
distribution of distances between wormhole entrances \cite{S15}. All such
facts enforce us think that wormholes should play the key role in explaining
the dark matter phenomenon.

However, there remains a strong scepticism in accepting the
existence of actual wormholes. It bases, in the first place, on
the fact that spherical
wormholes are highly unstable (they collapse during the characteristic time $%
\sim R/c$, where $R$ is the size of the throat section and $c$ is the speed
of light). Therefore, to be more or less stable (and traversable) they
require the presence of an exotic violating the averaged null energy
condition matter. It is possible to find a source of such matter at Planck
scales (e.g., due to the Casimir effect) but such a matter does not exist at
laboratory and astrophysical scales.

It turns out that the problem of the stability of cosmological
wormholes can be easily solved when we consider less symmetric
wormhole configurations \cite{S15}. In this case the presence of
an exotic matter is not the necessary condition for a wormhole to
be traversable in both directions and be stable, while the less
symmetry gives rise to the fact that cosmological wormholes have
the neck sections in the form of tori or even more complex
surfaces. To avoid misunderstanding we point out that under a
stable wormhole we do not mean static wormholes but rather
wormholes whose the characteristic time of evolution and the rate
of evolution are comparable with that of the Universe. In other
words, such wormholes do survive till the present days, though
they surely are not stationary objects. In the present paper we
explicitly demonstrate that a gas of wormholes can be obtained by
a specific factorization of the Lobachevsky space, or, in a more
general case, by the factorization of the Bianchi type -V
homogeneous spaces \cite{BKL}. Upon the factorization such spaces
lose the homogeneity property. Nevertheless they remain to be
locally homogeneous. This means that any sufficiently small
portion of such spaces (i.e., any point with its neighborhood)
coincides with analogous portion of the respective homogeneous
space.

\section{Open cosmological model}

First, we recall basic properties of the open Friedman models. Consider the
metric in the form%
\begin{equation}
ds^{2}=a^{2}(\tau )\left( d\tau ^{2}-\frac{4(d\vec{r})^{2}}{\left(
1-r^{2}\right) ^{2}}\right) .  \label{FM}
\end{equation}%
Here and below all symbols of vector operations (scalar product, etc) should
be interpreted in purely formal way as if the coordinates $\vec{r}=(x,y,z)$
were Cartesian. In particular, $(d\vec{r})^{2}=dx^{2}+dy^{2}+dz^{2}$ and $%
r^{2}=x^{2}+y^{2}+z^{2}$. The scale factor $a(\tau )$ is assumed to obey to
the standard Friedman equations without any exotic matter.

The space-like part of the above metric represents the Lobachevsky space
with the constant negative spatial curvature $k=-1$. In different
coordinates it takes the form%
\begin{equation}
dl^{2}=\frac{4(d\vec{r})^{2}}{\left( 1-r^{2}\right) ^{2}}=d\chi
^{2}+sh^{2}\chi \left( d\theta ^{2}+\sin ^{2}\theta d\phi ^{2}\right) .
\label{LM}
\end{equation}%
In the coordinates $\vec{r}$ \ the Lobachevsky space corresponds to the ball
$r^{2}\leq 1$ while the boundary sphere $r^{2}=1$ represents the absolute
(i.e., the spatial infinity). \ This space possesses the properties of
homogeneity and isotropy. The isotropy is straightforwardly seen from the
above form of the metric (it has the invariant form under rotations $%
x^{\alpha \prime }=U_{\beta }^{\alpha }x^{\beta }$, $U\in O(3)$),
while the homogeneity is not so explicit. It means that the metric
keeps the invariant
form with respect to translations $x^{\alpha \prime }=T^{\alpha }(\vec{r},%
\vec{r}_{0})$ where the vector $\vec{r}_{0}$ defines the value of the
displacement. We point out that in the present and next sections the
property of the isotropy of the space is not used and can be omitted.

Let $\vec{r}_{0}$ be an arbitrary point of the ball $r^{2}\leq 1$ and define
the translation which transforms this point to the origin $\vec{r}=0$. To
find the explicit law for the translation it is more convenient to use the
Poincare model of the Lobachevsky space in the form of upper half-space.
This is reached by introducing the Poincare variables (e.g., see \cite%
{KM95,KM97})%
\begin{equation}
\vec{\eta}+\vec{c}=\frac{2(\vec{r}+\vec{c})}{(\vec{r}+\vec{c})^{2}},
\label{PV}
\end{equation}%
where $\vec{c}$ is a point on the absolute. In terms of new coordinates the
metric of the Lobashevsky space becomes%
\begin{equation}
dl^{2}=\frac{\left( d\vec{\eta}\right) ^{2}}{\left( \vec{\eta}\cdot \vec{c}%
\right) ^{2}}  \label{LM2}
\end{equation}%
and the ball $r^{2}\leq 1$ transforms into the half-space $\left( \vec{\eta}%
\cdot \vec{c}\right) \geq 0$, while the absolute becomes the plane $\left(
\vec{\eta}\cdot \vec{c}\right) =0$. Indeed, the relation (\ref{PV}) can be
rewritten as
\begin{equation*}
\vec{r}=\frac{2\left( \vec{\eta}+\vec{c}\right) }{\left( \vec{\eta}+\vec{c}%
\right) ^{2}}-\vec{c}
\end{equation*}%
and therefore%
\begin{equation*}
1-r^{2}=\frac{4\left( \vec{\eta}\cdot \vec{c}\right) }{\left( \vec{\eta}+%
\vec{c}\right) ^{2}}\geq 0\text{.}
\end{equation*}

In terms of new coordinates it is easy to see the homogeneity of the
Lobachevsky space with respect to translations, i.e., that the metric (\ref%
{LM2}) does not change the form under the transformations%
\begin{equation*}
\vec{\eta}^{\prime }=\lambda \vec{\eta}+\vec{\xi}~,
\end{equation*}%
where $\lambda >0$ and $\vec{\xi}$ is the constant vector orthogonal to $%
\vec{c}$ (i.e., $\left( \vec{\xi}\cdot \vec{c}\right) =0$). The vector $\vec{%
\xi}$ describes translations along the directions orthogonal to $\vec{c}$,
while the constant $\lambda $ defines translations along the direction $\vec{%
c}$.

Let us take $\vec{c}=\vec{r}_{0}/r_{0}\,$. Then the new coordinates $\vec{r}%
^{\prime }$ (in which the point $\vec{r}_{0}$ is at the origin) are%
\begin{equation}
\vec{r}^{\prime }=T_{r_{0}}\left( \vec{r}\right) =\frac{2\left( \vec{\eta}%
/\eta _{0}+\vec{c}\right) }{\left( \vec{\eta}/\eta _{0}+\vec{c}\right) ^{2}}-%
\vec{c}  \label{Tr}
\end{equation}%
where $\eta _{0}=\frac{\left( 1-r_{0}\right) }{\left( 1+r_{0}\right) }$ and $%
\vec{\eta}(r)$ is given by (\ref{PV}). This transformation describes simply
the shift of space which reflects the homogeneity of space. In new
coordinates the point $\vec{r}_{0}$ transforms into the origin
\begin{equation*}
\vec{r}_{0}^{\prime }=\vec{r}^{\prime }\left( \vec{\eta}_{0}\right) =\vec{r}%
^{\prime }\left( \vec{r}_{0}\right) =0
\end{equation*}%
while the metric remains the same $dl^{2}=\frac{4(d\vec{r}^{\prime })^{2}}{%
\left( 1-r^{\prime 2}\right) ^{2}}$. We point out that translations have the
group properties which is known to correspond to the type-V homogeneous
space by the Bianchi classification \cite{BKL}.

\section{Wormhole connecting two Lobachevsky spaces}

The wormhole which connects two Lobachevsky spaces (or in a more
general case two Bianchi type-V homogeneous spaces) is obtained
from the Lobachevsky metric by a factorization, i.e., by imposing
periodic boundary conditions with respect to some part of
coordinates. Indeed, let us take an arbitrary point $O$ in the
space which we take as the origin of coordinates (this can be
reached by the shift transformation $T_{r_{O}}$ described in the
previous section) and issue a geodesic line from it, which we take
as the axis $OX$. Here we assume geodesics for the Lobachevsky
metric (one should not mix them with spacetime geodesics which are
trajectories of test particles in the total Friedman spacetime).
Now consider
two points on that line ($x=\pm a$) which are at the geodesic distance $%
d/2=\ln \frac{1+a}{1-a}=-\ln \eta _{a}$ from the origin and issue from them
a family of geodesics passing orthogonal to the axis $OX$. Such families of
geodesics form the 2-dimensional hyper-surfaces which represent Lobachevsky
planes and which are described by the equations%
\begin{equation}
\pm \frac{\left( \vec{r}\cdot \vec{a}\right) }{1+r^{2}}=\frac{a^{2}}{1+a^{2}}%
,  \label{HS}
\end{equation}%
where $\vec{a}=a\vec{\ell}$ and $\vec{\ell}$ $=(1,0,0)$ is the vector along $%
OX$ (or the point at which the geodesic $OX$ intersects the
absolute). In a flat space such surfaces would remain to be
parallel everywhere, while in the Lobachevsky space (of the
constant negative curvature) they start to diverge and represent
spheres which intersect the absolute under the right angles.
Now let us identify (glue) points on these two surfaces
and obtain a manifold which from the topological standpoint
corresponds to a cylinder. Then the geodesic $OX$ becomes the
shortest closed geodesic on such a manifold (which has the length
$d$).

It may be convenient to describe such a gluing in terms of new coordinates
which are introduced as follows (e.g., see appendix in \cite{KM95})%
\begin{equation}
\vec{u}=\frac{2\vec{r}}{1+r^{2}}.  \label{U}
\end{equation}%
In terms of new coordinates the absolute $r^{2}=1$ does not change its
position, i.e. $u^{2}=1$ and the Lobachevsky space remains to be the ball $%
u^{2}\leq 1$, while the above spheres become ordinary "flat" planes $%
u_{x}a=\left( \vec{u}\cdot \vec{a}\right) =$ $\pm
\frac{2a^{2}}{1+a^{2}}$. 

Since the Lobachevsky space is homogeneous, it is possible to continue the
cylinder to the whole Lobachevsky space. It is achieved by the translation
of the fundamental region $\ -\frac{2a}{\left( 1+a^{2}\right) }\leq
u_{x}\leq \frac{2a}{\left( 1+a^{2}\right) }$ along the axis $Ou$ on the
distance $d(a)=2\ln \frac{1+a}{1-a}$ (which is the shortest distance between
the two surfaces). Then the Lobachevsky space splits into "stripes" $%
u_{n}\leq u_{x}\leq u_{n+1}$, where $u_{n}=T_{a}^{2}(u_{n-1})$ is the
translation (which is defined by the vector $\vec{a}$, see below) and $%
n=0,\pm 1,\pm 2,...$($n\in
\mathbb{Z}
$) 
every of each coincides with the fundamental region (i.e.,
represents the same fundamental region). In other words, the cylinder
described above is the factorization of the Lobachevsky space over the
translation $u=T_{a}^{2}(u)$, while the Lobachevsky space itself represents
a cover of the fundamental region (of the cylinder).

The explicit law for such a translation is given by (\ref{Tr}). Indeed, let
us choose $\vec{c}=\vec{\ell}=(1,0,0)$, then in new coordinates the two
planes ($u_{x}=\pm \frac{2a}{\left( 1+a^{2}\right) }$) become the two
semi-spheres with the centers at the origin and radii
\begin{equation*}
R_{\pm }=\left( \frac{1-a}{1+a}\right) ^{\pm 1}.
\end{equation*}%
The plane $u_{x}=0$ (or $x=0$) in such coordinates becomes the semi-sphere
of the unit radius. Thus the translation along $\vec{\ell}$ \ which
transforms the spheres $R_{-}\rightarrow R_{+}$ is given by $\lambda $
\begin{equation*}
\lambda =\frac{R_{+}}{R_{-}}=\left( \frac{1-a}{1+a}\right) ^{2}
\end{equation*}%
and the system of stripes on the ball $u^2\leq 1$ 
becomes the system of rings restricted by
spheres of the radii%
\begin{equation*}
R_{n}=\lambda ^{n}R_{+}.
\end{equation*}%
This means that the cylinder is the factorization of the Lobachevsky space
with respect to re-scaling with $\lambda $, i.e. any two points related by
the relation $\vec{\eta}^{\prime }=\lambda ^{n}\vec{\eta}$ correspond to the
same point of the cylinder. In terms of initial coordinates the
factorization looks like%
\begin{equation}
\vec{r}\sim \vec{r}_{n}=T_{a}^{2n}(\vec{r})=\frac{2\left( \lambda ^{n}\vec{%
\eta}+\vec{\ell}\right) }{\left( \lambda ^{n}\vec{\eta}+\vec{\ell}\right)
^{2}}-\vec{\ell},  \label{T}
\end{equation}%
where the sign $\vec{r}\sim \vec{r}^{\prime }$ means that these two points
are equivalent (represent \ the same point) and
\begin{equation*}
\vec{\eta}=\frac{2\left( \vec{r}+\vec{\ell}\right) }{\left( \vec{r}+\vec{\ell%
}\right) ^{2}}-\vec{\ell}.
\end{equation*}

Thus, we see that the gluing of the two surfaces constructed does not change
the metric at all. It however changes properties of the geodesic lines and
matter fields defined on such a manifold. In particular, when we consider
density perturbations $\delta \rho (\vec{r},t)$ on such a background it
should obey the identity $\delta \rho (\vec{r},t)=\delta \rho (\vec{r}%
_{n},t) $ since $\vec{r}$ and $\vec{r}_{n}$ represent the same point of the
manifold. In all other respects the cylinder behaves like the standard
Lobachevsky space.

It however should be noted that upon the factorization the space
loses the property to be homogeneous. Indeed, the homogeneity
means that all points are equivalent (any point can be taken as
origin) and the space looks in the same way from any point of the
space. This surely does not retain upon the factorization. Indeed,
consider the geodesic $OX$ ($-a\leq x\leq a$) which goes through
the origin $O$ and represents the shortest closed geodesic. If we
take any other point $P$ of the space which does non lies on $OX$
axis and take it as origin, then we find that there are no closed
geodesics at all which go through the point $P$. In other words
such two points are not equivalent (the properties of space look
in a different way from these two points). Instead of global
homogeneity the factorized space possesses however the property of
the local homogeneity. This means that a sufficiently small
geodesic ball (in our case with geodesic distances less than
$d(a)$) around any point can be transformed (by an appropriate
translation) into the analogous ball around any other point $P$
(and it coincides with the analogous ball in the homogeneous
Lobachevsky space).

Consider now an orthogonal to \thinspace $OX$ geodesic issued from
the origin $O$ which without loss of generality we take as the
axis $OY$ and construct in the same way an additional couple of
hyper-surfaces which go through the points $y=\pm b$ and are
orthogonal to the axis $OY$. Now the shortest distance between
such surfaces is $d(b)=\ln \left( \frac{1+b}{1-b}\right)
^{2}=-\ln \lambda (b)$ and the surfaces are defined by the equations%
\begin{equation*}
\frac{2\left( \vec{r}\cdot \vec{b}\right) }{1+r^{2}}=\pm \frac{2b^{2}}{%
1+b^{2}}=bu_{y},
\end{equation*}%
where $\vec{b}=(0,b,0)$. Taking the point on the absolute
$\vec{c}=\vec{b}/b$ we define the analogous transformation
(\ref{T}) and an additional factorization of the Lobachevsky space
with the equivalence $\vec{r}$ $\sim $ $\vec{r}_{m}=$
$T_{b}^{2m}(\vec{r})$ ($m\in \mathbb{Z} $). The wormhole
configuration is the manifold which is obtained from the
Lobachevsky space by both factorizations, i.e., any two points are
equivalent (the same), if they are related by
\begin{equation*}
\vec{r}\sim \vec{r}_{n,m}=T_{b}^{2m}(T_{a}^{2n}(\vec{r}))
\end{equation*}%
for any $n$, $m\in \mathbb{Z}$. The two vectors $\vec{a}$ and
$\vec{b}$ are generators of such a factorization, while the point
$O$ defines the "central" point of the wormhole. We point out that
the section $z=0$ is the homogeneous space (torus of the constant
negative curvature) and all points on it can be taken as a
"central" point of the wormhole.

\section{Sections of the wormhole}

Now we are ready to describe the structure of sections of such a
wormhole. The simplest picture appears in coordinates (\ref{U}).
For any $u_{z}=const$ the section represents the torus, i.e., the
rectangle
\begin{equation}
-\frac{2a}{1+a^{2}}\leq u_{x}\leq \frac{2a}{1+a^{2}},\ \ -\frac{2b}{1+b^{2}}%
\leq u_{y}\leq \frac{2b}{1+b^{2}}  \label{Rc}
\end{equation}%
with opposite sides identified. Whether these inequalities
restrict a
finite region or not depends on the behavior of the absolute $\vec{u}^{2}=1$%
. On the $u_{x}$ -- $u_{y}$ plane the absolute (i.e., the infinity) is given
by $u_{x}^{2}+u_{y}^{2}=1-u_{z}^{2}$. If we assume $a^{2}+b^{2}<1$, then for
$u_{z}=0$ the torus (or the rectangle (\ref{Rc})) lies within the the
admissible domain $u_{x}^{2}+u_{y}^{2}\leq 1$ and the section is the torus
indeed. In particular, at $u_{z}=0$ such a torus has the smallest area. The
section in the form of the torus retains with the increase of $u_{z}^{2}$
till it reaches the value $\left( u_{z}^{0}\right) ^{2}=1-$ $\left( \frac{2a%
}{1+a^{2}}\right) ^{2}-\left( \frac{2b}{1+b^{2}}\right) ^{2}$. At
this value four points ($u_{x}=\pm \frac{2a}{1+a^{2}}$, $u_{y}=\pm
\frac{2b}{1+b^{2}}$ ) reach the absolute which means that they lie
at infinity and the torus starts to destroy. We point out that at
$u_{z}=\pm u_{z}^{0}$ the surface of
the torus still has a finite area. Let $a<b$, then for $u_{z}^{2}>u_{z}^{%
\ast 2}=1-\left( \frac{2a}{1+a^{2}}\right) ^{2}$ the absolute ($%
u_{x}^{2}+u_{y}^{2}=1-u_{z}^{2}$) gets within the rectangle (\ref{Rc}) which
means that the inequalities (\ref{Rc}) does not set any restriction on
admissible values of $u_{x}$ and $u_{y}$ and, therefore, the section is not
restricted. In other words, the sections $u_{z}=const$ represent the
sections of the unrestricted Lobachevsky space (without any factorization).

Thus, the two regions $u_{z}>u_{z}^{\ast }$ and
$u_{z}<-u_{z}^{\ast }$ correspond to two parts of the standard
Lobachevsky spaces which are connected only via the throat
($u_{z}^{2}<u_{z}^{\ast 2}$) in the form of a torus. In other
words such a manifold corresponds to a wormhole configuration
which connects two Lobachevsky spaces (we recall that any section
$u_{z}=const$ divides the Lobachevsky space on two equal parts).
If we identify these two spaces with the help of some of motions
of the Lobachevsky space, we may obtain a wormhole which connects
two different regions in the same Lobachevsky space which we
describe below.

\section{Handle type wormhole}

The handle type wormhole configuration is constructed from the
above construction as follows. First, consider a point $O_{1}$
with coordinates being $\vec{r}_{1}$ and two generators
$\vec{a}_{1}$ and $\vec{b}_{1}$ (for simplicity we assume $\left(
\vec{a}\cdot \vec{b}\right) =0$, though this is not a necessary
condition) which define the factorization of the Lobachevsky space
to the wormhole configuration described in the previous section.
In the explicit form the factorization is given by the the
equivalence
relations $\vec{r}\sim \vec{r}^{\prime }=\mathbf{T}_{A}^{nm}(\vec{r}%
)=T_{r_{1}}^{-1}T_{a_{1}}^{2n}T_{b_{1}}^{2m}T_{r_{1}}\left( \vec{r}\right) $%
, where $n$, $m\in
\mathbb{Z}
$, $T_{r_{1}}$ defines the shift of the origin to the point $O_{1}$, and $%
A_{1}$ defines the set of parameters $A_{1}=(\vec{r}_{1},\vec{a}_{1},\vec{b}%
_{1})$. Upon the factorization the Lobachevsky space reduces to the wormhole
described in the previous section. The point $O_{1}$ defines the central
point of such a wormhole $\vec{r}_{1}^{\prime }=T_{r_{1}}\left( \vec{r}%
_{1}\right) =0$, while the regions $z_{1}>0$ and $z_{1}<0$ can be considered
as two Lobachevsky spaces from which we cut a solid torus. Consider now an
additional point $O_{2}$ (let it be in the region $z_{1}>0$ and lie outside
the solid torus $u_{z_{1}}>u_{z_{1}}^{\ast }$) and define another two
generators $\vec{a}_{2}$ and $\vec{b}_{2}$ such that $a_{1}=\left\vert \vec{a%
}_{1}\right\vert =a_{2}$ and $b_{1}=b_{2}$. Then they define another
factorization and an additional wormhole which also separates the upper part
of the Lobachevsky space $z_{1}>0$ into two parts $z_{2}>0$ and $z_{2}<0$,
which are also the two Lobachevsky spaces with a some region (in the form of
solid torus) being removed. The relations $a_{1}=a_{2}$ and $b_{1}=b_{2}$
insure that such solid tori are equal and, therefore, we can match and
identify the regions $z_{1}<0$ and $z_{2}>0$. To match these two regions we
have to use the composition of the translation from the point $O_{1}$ to the
point $O_{2}$ (which is given by the map $T_{1\rightarrow
2}=T_{r_{2}}^{-1}T_{r_{1}}$) and the rotation $U$ which matches the axis $%
O_{2}Z_{2}$ ($z_{2}>0$) and $-O_{1}Z_{1}$ ($z_{1}<0$). This assumes an
additional factorization of the initial \ Lobachevsky space by the
equivalence relation $\vec{r}\sim \vec{r}^{\prime }=UT_{r_{2}}^{-1}T_{r_{1}}(%
\vec{r})$. The resulting manifold corresponds to the wormhole configuration
which connects (through the handle in the form of a torus) two regions in
the same space. From the visual standpoint such a space corresponds to a
Lobachevsky space in which we cut two equal solid tori and glue them by the
surfaces of the tori.

One particular degenerate configuration of such a wormhole serves of a
special interest. It is the so-called string -like configuration. Indeed, it
realizes in the situation when both points $O_{1}$ and $O_{2}$ belong to the
same plane $z_{1}=z_{2}=0$. In this case the additional factorization
reduces to the reflection $z_{1}=-z_{2}$ and a shift on the two-dimensional
plane $z=0$ ($T_{1\rightarrow 2}$). From visual standpoint in this case
instead of two solid tori we will get a single torus (positions of the two
tori merely coincide), while such a configuration corresponds to a closed
cosmic string. In the case when $b\rightarrow 1$ one of the torus radii
becomes infinite and such a wormhole becomes an open cosmic string.

\section{Wormhole in a conformaly flat spacetime}

It may be convenient to consider wormholes in a flat spacetime.
Let us
change the coordinates in the interval (\ref{FM}) according to%
\begin{equation*}
x^{0}=e^{\tau }\frac{1+r^{2}}{1-r^{2}},\ \vec{x}=2e^{\tau }\frac{\vec{r}}{%
1-r^{2}}
\end{equation*}%
then the ordinary flat interval reduces to
\begin{equation*}
ds^{2}=\left( dx^{0}\right) ^{2}-\left( d\vec{x}\right) ^{2}=e^{2\tau
}\left( d\tau ^{2}-4\left( \frac{d\vec{r}}{1-r^{2}}\right) ^{2}\right)
\end{equation*}%
therefore from (\ref{FM}) we find%
\begin{equation*}
ds^{2}=a^{2}(\tau )e^{-2\tau }\left( \left( dx^{0}\right) ^{2}-\left( d\vec{x%
}\right) ^{2}\right)
\end{equation*}%
where the variable $\tau $ is given by
\begin{equation*}
e^{2\tau }=\left( x^{0}\right) ^{2}-\left( \vec{x}\right) ^{2}
\end{equation*}%
and therefore, such coordinates do not cover the whole spacetime.
In terms of new coordinates the hyper-surfaces (\ref{HS}) become
flat but are depending on time
\begin{equation}
\left( \vec{x}\cdot \vec{a}\right) =\pm \frac{2a^{2}}{1+a^{2}}x^{0}.
\end{equation}%
They move in space with the velocities%
\begin{equation*}
\vec{V}=\pm \frac{2\vec{a}}{1+a^{2}}.
\end{equation*}%
Since $a^{2}\leq 1$, such velocities are always less than the
speed of light ($V^{2}\leq 1$) and can be very small (for
$a^{2}\ll 1$, $V\simeq 2a$). Thus in these coordinates the
space-like part will represent the standard flat torus whose both
radii expand with the speed proportional to their size. The bigger
radius the bigger speed of expansion. We stress that such boundary
conditions have the direct analogy to moving flat mirrors in the
flat spacetime which were used in describing the general behavior
of the metric near the cosmological singularity in classical and
quantum gravity, e.g., see Refs. \cite{K92} and references
therein.

\section{Dark matter phenomenon}

When the gravitational field is rather weak, the time component of
the Einstein equations for scalar perturbations reduce to the
standard Newton's equation \cite{Peeb}
\begin{equation}
\delta R_{00}\simeq\frac{1}{a^{2}}\Delta \phi =4\pi G\left( \delta
\rho +\frac{3}{c^{2}}\delta p\right) , \label{PT}
\end{equation}%
here $a$ is the scale factor of the Universe, $R_{00}$ is the time component of the Ricci tensor,
$\phi $, $\delta \rho $ and $%
\delta p$ are the scalar metric, mass density and pressure
perturbations
respectively. The Laplace operator $\Delta \phi =\frac{1}{\sqrt{\gamma }}%
\partial _{\alpha }\left( \sqrt{\gamma }\gamma ^{\alpha \beta }\partial
_{\beta }\phi \right) $ is constructed with the help of the metric (\ref{LM}%
). The presence of a factorization (of wormholes) changes properties of
perturbations defined on such a manifold. In particular, the density
perturbations $\delta \rho (\vec{r},t)$ should obey the identity $\delta
\rho (\vec{r},t)=\delta \rho (\vec{r}^{\prime },t)$ for any two points $\vec{%
r}\sim \vec{r}^{\prime }=T(\vec{r})$ related by one of possible
equivalence relations (which means periodicity with respect to
some appropriate coordinates). Therefore, the behavior of
perturbations can be determined by
the Green function which corresponds to a unit source%
\begin{equation}
\Delta G(x,x^{\prime })=\frac{4\pi }{\sqrt{\gamma }}\delta (r-r^{\prime }).
\label{gf}
\end{equation}%
First, let us set the position of the source $r^{\prime }$ to the origin
(which is made by translation $T_{r^{\prime }}$ defined by (\ref{Tr})). Then
the Green function (i.e., the Newton's law for the Lobachevsky space) looks
like%
\begin{equation*}
G(r)=-\frac{1+r^{2}}{r}.
\end{equation*}%
For $r\ll 1$ it gives the standard Newton's law $G\simeq -1/r$. Now making
use the back transformation $T_{r^{\prime }}^{-1}$ we find%
\begin{equation*}
G(r,r^{\prime })=-\frac{1+\left( T_{r^{\prime }}^{-1}(\vec{r})\right) ^{2}}{%
\left\vert T_{r^{\prime }}^{-1}(\vec{r})\right\vert }.
\end{equation*}

The presence of a gas of wormhole means the existence of the factorization
of space. This means that all points which are connected by transformations
of the type $r_{N,k}^{\prime }=\mathbf{T}_{A_{k}}^{n_{k}m_{k}}\left( \vec{r}%
^{\prime }\right) $ represent the same point of space (where $N$
numerates different wormholes and $k$ all parameters $A_{k}$,
$n_{k}$, and $m_{k}$ of a particular wormhole). Therefore, in
terms of the Lobachevsky space (which represents a cover for the
actual physical space) the unit source multiplies. It acquires an
infinite number of images, while the right-hand side of (\ref{gf})
transforms into
\begin{equation*}
\frac{1}{\sqrt{\gamma }}\delta (r-r^{\prime })\rightarrow \frac{1}{\sqrt{%
\gamma }}\sum_{N,k}\delta (r-r_{N,k}^{\prime })=\frac{1}{\sqrt{\gamma }}%
\delta (r-r^{\prime })+b(r,r^{\prime }).
\end{equation*}%
This defines a halo of "dark matter" around any point source
$b(r,r^{\prime })$. The structure of such a halo is defined by the
distribution of
wormholes (i.e., by distributions of points $O_{k}$ and generators $\vec{b}%
_{k},\vec{a}_{k}$). From the formal standpoint the equation for
perturbations (\ref{PT}) remains correct (as the microscopic
equation), but it requires very complex boundary conditions (on
wormhole throats, or due to the factorization of the background
space). Exactly like in macroscopic Electrodynamics it is more
convenient to introduce the specific topological permeability and
describe effects of wormholes (of the factorization) as a
modification of the perturbation theory. In linear case, if we
neglect possible peculiar motions of wormholes, it is given by the
bias function \cite{KT} which transforms the right hand side of
(\ref{PT}) according to
\begin{equation}
 \delta \rho (r) =\delta \rho (r) +\int b(r,r^{\prime })\delta \rho (r^{\prime
 }) d^3r^{\prime} .
 \label{BM}
 \end{equation}%
 In a more general case the bias function depends on time which
 more essentially modifies the perturbation theory.
We do not discuss this problem here in more details, see however
the more substantial discussions of this problem in Refs.
\cite{KS11,KS07,KT}.

\section{Discussions}
In the present paper we described in details the factorization of
the Lobachevsky space to a constant negative curvature space which
contains an arbitrary collection of wormholes. Such a space
evolves as the open Friedmann model, while the presence of a gas
of wormholes does not require any form of exotic matter. It is
straightforward to generalize such a factorization on the case of
a homogeneous Bianchi type- V homogeneous spaces, however, in this
case the handle type wormhole will add some restrictions on
possible orientations of the conjugated throat entrances.

There remains an important problem to study all possible
factorizations of the rest Bianchi types of homogeneous spaces and
verify which of them may correspond to non-trivial topological
(wormhole-like or other) configurations. We point out that in
general, upon the factorization the homogeneous space loses the
property of global homogeneity, though the local homogeneity
retains. The exclusion is the case when the factor group $G/H$
possesses the group properties, i.e. represents a group. Here $G$
is the group of homogeneity (group of translations) and $H$ is a
finite subgroup (e.g., the subgroup defined by a translation $T_a$
on the distance $a$ and all its degrees $T_a^n$, $n\in \mathbb{Z}$
which define a discrete subgroup of the group of translations). In
this case the space remains to be globally homogeneous. The
simplest example is the Bianchi type - I models. For this spaces
translations is the standard shift on a constant vector as in the
standard Euclidean space. The factorization with respect to a
specific translation (displacement on a constant vector)
transforms the Euclidean space in the ordinary cylinder (periodic
in the direction of the above displacement). The cylinder also
represents a homogeneous space, since in contrast to the wormhole
in the Lobachevsky space, it does not contain distinguished
points, while the factor group $G/H$ represents a new group which
acts on the cylinder. In this case the Bianchi type -I spaces
represent the universal covering for the cylinder.

In this manner, we see that stable cosmological wormholes may
indeed exist without any exotic matter and they may survive till
the present days. A stable cosmological wormhole has the neck
section in the form of a torus and, therefore, it may leave
specific imprints in CMB in a form of rings. This may explain the
results of Ref. \cite{MNR} where the presence of such ring-type
structures in CMB was established. See also discussion of this
problem in Ref. \cite{KS15}.

In confronting to observations there remains a strong problem
though. Indeed, observations of $\Delta T/T$ spectrum indicate
that the spatial curvature of our space is very close to zero,
while the construction of a stable wormhole suggested seems to
require the presence of the negative curvature.

To overcome this problem we see two possible  ways. First one is
to introduce inhomogeneities. Indeed, the factorized Lobachevsky
space is already an inhomogeneous space (though it may have the
property of local homogeneity). In other words, every wormhole can
be considered as an inhomogeneous structure in the background
space. Therefore, in principle, we may add some amount of matter
density outside of the wormhole necks to make the background space
close to the flat space (the matter density should be closer to
the critical value). However it is not possible to do for throats
themselves which means that throats would expand a little bit
faster than the background space. While the typical size of
wormhole throats is much smaller, than the Hubble radius, this
will not destroy the observed mean homogeneity and isotropy of the
Universe. We point out that from the astrophysical standpoint
wormhole throats look like specific compact astrophysical objects
and the fact that they introduce some local inhomogeneity is not
surprising. The second way is to study more complex wormhole-type
configurations.

There also remains the problem to consider the modification of the
perturbation theory in the presence of wormholes. From the
microscopic (or formal) standpoint the Lifshitz theory \cite{L46}
(see also \cite{Peeb}) does not change at all, since microscopic
equations have local character and remain the same. However they
require too complex boundary conditions (due to the factorization
of the background space). The best way to account for the presence
of wormholes is to introduce in the equations the permeability of
space which have the topological origin. For example, the presence
of wormholes may be described phenomenologically as the bias
(\ref{BM}) which can be interpreted as the presence of the dark
matter. This essentially simplifies the consideration of the
behavior of perturbations. During the development of metric
perturbations wormholes start to move and the total picture
becomes very complex and even nonlinear.   Some steps in this
direction we did in Ref. \cite{KS11} but those are clearly not
sufficient.

\end{document}